\documentclass[journal]{IEEEtran}

\usepackage{epsfig}
\usepackage[linesnumbered,ruled,vlined]{algorithm2e}

\usepackage{graphics}
\usepackage{graphicx} 
\begin{document}
\title { \Huge QoS Group Based Optimal Retransmission Medium Access Protocol for Wireless Sensor Networks}
\author{Kumaraswamy M*, Shaila K*, Tejaswi V*, Venugopal K R*, S S Iyengar** and L M Patnaik*** \\
        ** Department of Computer Science and Engineering,\\
          University Visvesvaraya College of Engineering, Bangalore University, Bangalore 560 001 \\  
        *** Director and Ryder Professor, Florida International University, USA, \\
              **** Honorary Professor, Indian Institute of Science, Bangalore, India \\
                  kumarasm67@yahoo.co.in\\}

\maketitle
\begin{abstract}
This paper presents, a Group Based Optimal Retransmission Medium Access (GORMA) Protocol is designed, that combines protocol of Collision Avoidance (CA) and energy management for low-cost, short-range, low-data rate and low-energy sensor nodes applications in environment monitoring, agriculture, industrial plants etc. In this paper, the GORMA protocol focuses on efficient MAC protocol to provide autonomous Quality of Service (QoS) to the sensor nodes in one-hop QoS retransmission group and two QoS groups in WSNs where the source nodes do not have  receiver circuits. Hence, they can only transmit data to a sink node, but cannot receive acknowledgement control signals from the sink node, which improves transmission reliability in WSNs. The proposed protocol GORMA provides QoS to the nodes which work independently on predefined time by allowing them to transmit each packet an optimal number of times within a given period. Our simulation results demonstrate the performance of GORMA protocol which maximize the delivery probability of one-hop QoS group and two QoS groups and minimize the energy consumption.

\end{abstract}


\section{INTRODUCTION}
Wireless Sensor Networks \textit{(WSNs)} consists of a several autonomous low-cost sensor nodes and distributed over a wireless networks. The sensor nodes architecture includes a microprocessor, sensor, actuator devices and a radio communication device. \textit{WSNs} provide a wide range of potential applications like healthcare, environmental monitoring, battlefield monitoring, remote sensing, industrial process control, surveillance and security etc. A typical \textit{WSNs} consists of one or more sink nodes and large number of data sensor nodes deployed, each source nodes generates data and transmits to the sink through common radio communication module.  

In general, such networks consists of both transmitting and receiving circuits. However, in most applications the devices generally collect data and transmit to sink node. The communication from sink node to source node is minimum. So, the receiver circuitry adds extra cost and also consumes significant amount of energy during process. Thus by using sensor data nodes with only transmitters, the device cost and the entire network infrastructure cost can be reduced. These wireless sensor devices are equipped with sensing, computation and wireless communication capabilities. Sensing tasks for sensors devices could be temperature, light, sound, humidity, vibration, etc. The Direct access scheme \textit{WSNs} communication devices are low rate communication protocol are designed for low cost, low data rate and low power \textit{WSNs} devices. 

Optimal retransmission is the process of sending packets to the sink multiple number of times to achieve the maximum delivery probability. The optimal retransmission in \textit{WSNs} is mainly focused on \textit{QoS} in terms of packet delivery probability and energy efficient. 

In this paper, we consider \textit{WSNs} that deals with \textit{QoS} group based medium access control scheme of different groups that has low complexity, less power consumption and optimum cost. Proposed scheme consists of network topology with more number of source nodes which are distributed and decentralized in one-hop communication range. In present day environment every source node in a \textit{WSNs}is equipped with only transmitter module. The receiver module is avoided, since they consume more energy and are expensive due to the hardware complexity. The throughput requirement are low because the source collects and transmits the data to the sink. The sink node in the network has both transmitter and receiver and it receives the data transmitted by the source nodes. There are large number of applications which use the above concept such as Smart Environment \cite{IAE}, Structure Monitoring \cite{NXU}, Smart Home Monitoring\cite{DCB}, Green House Monitoring \cite{DSM}, Intelligent Transportation \cite{AVN}, Smart Kindergarten \cite{CMY} and Medical Monitoring \cite{AWL}.
   
Most of the medium access control protocol are like polling, CSMA \cite{LQW}, Automatic Repeat  Request (ARQ), collision avoidance/detection \cite{JBT} and scheduled transmissions \cite{YHE} are not effective because they need the \textit{ACK} to transmit the next packet. 

\noindent\textit{Motivation}\\
In many application scenarios of sensor networks, sensor data must be delivered to the sink node within time constraints. It is crucial to evaluate the performance limits, such as maximum data delivery and energy consumption of traffic loads under all conditions. 

Moreover, sensor networks present several technical challenges in terms of extremely low cost, low energy requirements and limited communication capabilities, while dealing with various workloads and diverse constraints. Addressing all these problems inevitably requires performance analysis techniques to provide insight on the design parameters and system behavior.\\

Hence, optimal retransmission, maximization of the packet delivery probability and energy efficiency have to be consider in desiging \textit{WSNs}.

\noindent\textit{Contribution} \\
This paper presents, a Group Based Optimal Retransmission Medium Access Protocol (\textit{GORMA}) provides \textit{QoS} to the sensor nodes using direct access mode where each node transmits data packet by selecting variable slot randomly for adaptive data packet considering local environment. Nodes can only join the network during direct access periods. The time interval between direct access periods could be small. So, in the proposed protocol, nodes randomly decide whether it should retransmit to help the packet delivery depending on some pre-calculated optimal retransmission probabilities. The sink node receive exactly one error-free retransmission data packet without collision. 
 
The main goal of this paper is to find a solution for  optimal retransmission that every single node should try for each of the data packet to achieve their required \textit{QoS} in terms of high probability of data delivery. An analytical method to evaluate the maximum data delivery probability and minimum energy consumption is proposed. First, we design a Group Optimal Retransmission Medium Access Protocol (\textit{GORMA}),  which is simple and compatible with the 802.11b standard to provide maximum data delivery.  \\

\noindent\textit{Organization}\\
The rest of the paper is structured as follows. Related works are discussed briefly in Section II. An overview of the Background is given in Section III. In Section IV we describe the Problem Definition and provide the required objective and assumptions. In Section V we introduce the System Model. In Section VI Mathematical Model discuses the One-Hop Retransmission and Two-QoS Groups. The proposed Algorithm discuses in Section VII. In Section VIII Performance Evaluation present the simulation result. Finally, Conclusions in Section IX.\\

\section{RELATED WORK}

Das et al., \cite{DCB} to developed a intelligent home which acts as an intelligent agent. The main objective of this agent is to maximize the comfort of inhabitant who is using it and minimize the cost of this operation. We can control each and every device in the intelligent home like during dawn the MavHome switches on the heat adjusts the temperature to optimal for waking up. Stipanicev et al., \cite{DSM} propose embedded systems that are more popular nowadays in monitoring and controlling of objects which are far and disjoint. 

Andrisano et al., \cite{AVN} designed an intelligent transport system in 3G radio networks. That rely on active sensors to provides live transmission of data. It overcomes the limitation in autonomous systems and concentrates on the drivers safety. Chen et al., \cite{CMY} propose the smart kindergarten infrastructure which consists of sensors, embeded in physical objects. This creates a smart environment called as SmartKG is implemented in kindergarten where the sensor is deployed to capture the interactions between students, teachers and the interaction objects like toys. 

Anliker et al., \cite{AWL} developed a system that deals with the human health. \textit{AMON} is wireless network which is used for alerting human cardiac/respiratory system in case of  heart patients. This system collects and evaluates the signals detected by intelligent multiparameter medical emergency detector which inturn has cellular connection to one of the medical center. 

Pai et al., \cite{PSH} designed a novel adaptive retransmission algorithm to improve the misclassification probability of distributed detection with error-correcting codes in fault-tolerant classification system for Wireless Sensor Netorks. The local decision of each sensor is based on its detection result. The detection result must be transmitted to a fusion center to make a final decision. 

Lu et al., \cite{LFQ} proposed a \textit{MAC} layer cooperative retransmission mechanism and a node can retransmit lost packets on behalf of its neighboring node. However, although each lost packet can be recovered by a neighboring node, it still requires a new transmission for each retransmission attempt, which largely limits its ability to increase the throughput of the network. 

Cerutti et al., \cite{CFH} proposes fixed Time-Division Multiple-Access \textit{(TDMA)} scheme delivery. When a node overhears a neighbours unsuccessful packet, it may retransmit that packet in its own allocated slot, provided the queue of its own packets is empty. Dianati et al., \cite{DLN} presents concurrent cooperation communication among the nodes to retransmit a packet to the destination if they receive the corresponding negative acknowledgement from the destination.

Xiong et al., \cite{XLG} considers cooperative forwarding in \textit{WSNs} from a \textit{MAC}-layer perspective, which means a receiver can only decode one transmission at a time. Fan et al., \cite{FLS} propose an interesting \textit{MAC} layer anycasting mechanism and randomized waiting at the application layer, to facilitate data aggregation spatially and temporally in structure-free sensor networks. They address the collision problem by proposing a modified \textit{CSMA/CA} protocol and randomized waiting scheme to reduce the number of retransmissions.

Noh et al., \cite{NES} propose Active Caching \textit{(AC)} to achieve desired Communication Reliability \textit{(CR)} levels of the various sensor network applications. This is a flexible loss recovery mechanism, when the packet delivery rate during multi-hop transmission from a source to an intermediate node decreases below the \textit{CR}, \textit{AC} retransmits lost packets from the source to the intermediate node so that the intermediate node has all data packets just like the source node.

Qureshi et al., \cite{QCJ} propose a latency and bandwidth efficient coding algorithm based on the principles of network coding for retransmitting lost packets in a single-hop wireless multicast network and demonstrate its effectiveness over previously proposed network coding based retransmission algorithms.

Aggelos et al., \cite{AKD} present a transmit time-offset based distributed relay selection technique, none of the nodes takes the leading role to decide the best relay, instead each relay node assigns itself a transmit time-offset. Each cooperating relay is in receive-mode or carrier sense  mode to detect other relays transmission while timer is counted down. The relay node with the best channel quality has smallest timer value and it transmits first among all the relays. As it transmits, the other relays hear the transmission and stop their competition to transmit.

He et al., \cite{HLF} propose the single-relay Cooperative Automatic Repeat Request \textit{(CARQ)} protocol. In \textit{CARQ} best relay node is selected in a distributed manner by relays using different backoff time before packet retransmission. In a dense network, due to high possible collision probability among different contending relays, an optimized relay selection scheme is introduced  to maximize system energy efficiency by reducing collision probability. 

Chu et al., \cite{CAE} present relay selection and selection diversity for coded cooperation in Wireless Sensor Networks, with complexity attribute for the sensor nodes. In earlier methods, a relaying technique based on Repeat-Accumulate \textit{(RA)} codes was introduced, where it was imagined that the relay does not carry out decoding and simply uses demodulated bits to form codewords.

Suriyachai et al., \cite{SRS} provides \textit{QoS} support by giving deterministic bounds for node-to-node delay and reliability. It can be a suitable for applications requiring absolute delay and reliability assurance. The collision-free \textit{TDMA} scheme are divided into fixed-length portions called epochs. In each epoch, a sensor node has \textit{k} exclusive slots for only single \textit{DATA-ACK} message exchange. If a sensor node does not have any data to send, it sends a simple control message at the first reserved slot indicating that it will not send anything in this epoch. Energy consumption is reduced by employing different duty cycles for each sensor node depending on their number of child nodes in the predetermined data gathering tree.

Ruiz et al., \cite{RVM} propose an architecture collaboration in which the \textit{MAC} and routing protocols to discover and reserve routes to organize nodes into clusters and to schedule the access to the transmission medium in a coordinated time-shared fashion. It achieves \textit{QoS} and reduces energy consumption by avoiding collisions and considerably lowering idle listening.

Tannious et al., \cite{TAN} present the secondary node user exploits the retransmissions of primary node user packets in order to achieve a higher transmission rate. The secondary node receiver can potentially decode the primary node users packet in the first transmission and opportunities are properly exploited. 

Bai et al., \cite{BEX} propose a design of \textit{IEEE} 802.11 based wireless network for \textit{MAC} that dynamically adjusts the retransmission limit to track the optimal trade-off between transmission delay and packet losses to optimize the overall network control system performance. 

Volkhausen et al., \cite{VDK} focuses on cooperative relaying, it exploits temporal and spatial diversity by additionally transmitting via a relay node, such relaying improves packet error rates and transmit only once rather than on each individual hop along the routing path. This cooperation reduces the total number of transmissions and improves overall performance.

Levorato et al., \cite{LMZ} propose the optimal throughput is achieved by the secondary node users in wireless networks when the primary node user adopts a retransmission-based error control scheme. The secondary node users maximize the throughput, with a constraint on the performance loss and an increased failure probability of the primary node user. 

Wang et al., \cite{WLC} propose the local cooperative relay for opportunistic data forwarding in mobile ad-hoc networks. The local cooperative relay select the best local relay node without additional overhead, such real time selection can effectively bridge the broken links in mobile networks and maintain a robust topology.

\section{BACKGROUND}  
Sudhaakar et al., \cite{RSJ} propose a novel Medium Access Control scheme,  considering a new class of low complexity, low power and low-cost wireless networks. Such  network typically consists of large number of source nodes, which are within one-hop communication range to one or few sink nodes. Each of these source nodes is equipped with only transmitter module in order to eliminate the cost due to hardware complexity and energy consumption of the receiver module. As a result, they are not capable of receiving any signals like \textit{ACK/NAK}. The source nodes collect data and transmit relatively small data frame to the sink nodes once a while and hence the throughput requirement of the source nodes is low. The sink nodes are the only nodes in the network that are equipped with receiver modules and are capable of receiving the transmissions of the source nodes.

\section{PROBLEM DEFINITIONS}
Consider a Wireless Sensor Network consisting of \textit{N} nodes as shown in Figure 1, having source nodes and one or two sink nodes. All the \textit{WSNs} nodes are within one hop transmission range of the sink. The source nodes do not have receiver unit, so it is impossible for sensor nodes to sense the channel for collision detection or receive any acknowledgements from the sink node. The main objective of the proposed work is to  

\begin{figure}
\begin{center} 
\includegraphics[width=20pc,height=10pc]{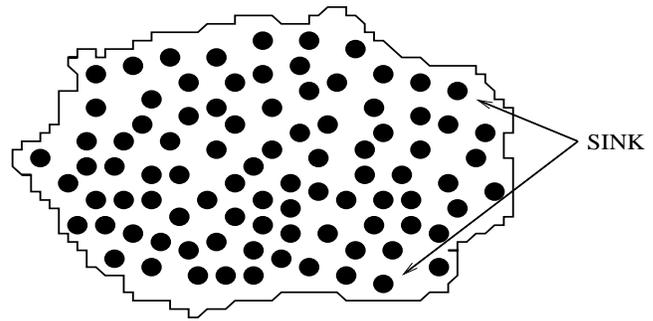}
\caption{Nodes Deployment in Wireless Sensor Network}
\end{center}
\end{figure}

\begin{enumerate}
\item 
Maximize the delivery probability of the nodes in the \textit{QoS} groups. 
\item 
The number of retransmissions in the network is minimized, this reduces energy consumption of sensor nodes and maximizes the packet delivery probability.
\item
Sink node in the network guarantees to receive the sent packet with probability.
\end{enumerate}

\subsection{Assumptions}
\renewcommand{\labelenumi}{(\roman{enumi})}
\begin{enumerate}
\item All nodes are homogeneous.
\item The nodes are randomly distributed within a area.
\item Packet generation rate at each node follows a Poisson distribution.
\item Only one wide channel is available for all communicating nodes.
\end{enumerate}

\section{SYSTEM MODEL}

The \textit{GORMA} provides \textit{QoS} to the nodes using direct access mode where each node independently depending on local conditions to transmits data packet by selecting variable slot randomly for variable data packet according to the number of sensors nodes. Nodes can only join the network during direct access periods. The time interval between direct access periods could be small. In the proposed protocol \textit{GORMA}, nodes randomly decide whether it should retransmit to help the packet delivery depending on some pre-defined optimal retransmission probabilities. The sink node receive exactly one error-free transmission packet in a slot, without collision with other simultaneous transmissions. Our goal is to develop decentralized \textit{MAC} protocol to provide \textit{QoS} guarantees for both time-critical and non time-critical sensor applications. This is a challenge that has not been addressed by any existing approach. The most important metrics to analyze the \textit{QoS} performance of \textit{MAC} protocol is packet delivery probability and energy efficiency. 
 
In \textit{WSNs}, each sensor node data packets transmission duration is relatively small when compared to the data packets that are generated at a constant rate \textit{i.e}, one packet every \textit{T} units of time. In addition, if a packet cannot be successfully delivered within a data generation \textit{T} units of time, the data packets is simply neglected. This makes sures that the new data packets have greater chance of being successfully delivered. Thus the maximum delivery probability that can be achieved by each individual sensor node increases eventually, so that all the nodes in the network achieve their required \textit{QoS} in terms of data delivery probability. 

In addition, we consider a network architecture in which the source nodes generate data and transmit periodically to the sink node. The set of \textit{N} nodes are partitioned into two \textit{QoS} groups $C_{1}$ and $C_{2}$, with each group containing $m_{1}$ and $m_{2}$ nodes respectively. The packet transmission duration $T_{p}$ of all nodes are assumed to be same. Each node in $C_{1}$ and $C_{2}$ requires minimum packet delivery probability $d_{k}$ (1 $\leq$ j $\leq$ 2). The protocol \textit{GORMA} is to find the efficient number of retransmissions $y_{k}$ for each $C_{k}$, such that, if every node in $C_{k}$ transmits $y_{k}$ times in every $T_{p}$ units of time, it should achieve a delivery probability of at least $d_{k}$. 

\begin{table}
\begin{center}
\caption{\bf Notations}
\renewcommand{\tabcolsep}{0.8pc}
\renewcommand{\arraystretch}{1.4}
\begin{center}
\begin{tabular}{|l|l|}
\hline 
  Symbols  &  Meaning \\
\hline
N           & Total number of sensor nodes  \\
\hline
T           & Data Packet Generation time   \\
\hline
$C_k$       & Number of QoS Group \\ 
\hline
$m_k$       & Number of nodes in each group \\
\hline
$T_p$       & Duration of packet transmission \\
\hline
$d_k$       & Minimum packet delivery probability \\
\hline
$y_k$       & Number of retransmission  \\
\hline
$\beta$     & Packet arrival rate  \\
\hline
p           & Number of packets  \\
\hline
Q           & Packet delivery probability  \\
\hline
$q_k$       & Minimum delivery probability \\
\hline
$\beta_{tg}$ & Overall traffic generated by all nodes \\
\hline      
$T_t$       & Time of $p$ packet transmission  \\
\hline
$\tau_{cs}$ & Carrier sense period of packet \\
\hline
$t_k$       & Time of retransmission in $C_k$  \\
\hline
\end{tabular}
\end{center}
\end{center}
\end{table}

\section{MATHEMATICAL MODEL}

\subsection{ONE-HOP RETRANSMISSION}
We have assumed that the source nodes generate data at constant rate of one packet every \textit{T} units of time and the retransmission time for each packet is much smaller than the duration of packet transmission \textit{$T_{p}$}. To achieve equal packet delivery probability by all the nodes in the \textit{WSNs}. Under this assumption the packet arrival rate can be modeled as a Poisson distribution. The number of nodes in the network is denoted by \textit{N} and the number of retransmissions by each node for each packet is denoted by \textit{$y_{k}$}. The notations are defined in Table I.

The packet arrival rate of the source nodes can be modeled as a Poisson distribution and the probability that \textit{p} packets are transmitted in an interval \textit{$T_t$} with \textit{Q(N)} the probability of \textit{N} arrivals in one time slot is given by  

\begin{equation}
Q(N) = \frac{(\beta T_t)^{p}}{p!}e^{-\beta T_t}  
\end{equation}

Where, $\beta$ is the rate of traffic generated by all other nodes inside the transmission range of a node and is equal to $\frac{(N-1)}{T}y$. 

The probability that the packet transmitted by node \textit{k} does not collide, so it is same as the probability that no packet were transmitted by the other \textit{N-1} nodes in an interval 2$T_{p}$. Therefore $Q_{nc}$ is  

\begin{equation}
Q_{nc} = e^{-2\beta T_{p}}
\end{equation}

The above discussion presents the probability that a packet transmitted by node \textit{k} is successfully received by the sink. However, node \textit{k} transmits \textit{$y_{k}$} copies of the packet at random instants in every time interval \textit{$T_{p}$}. Hence the actual parameter of interest will be the probability that at least one of these \textit{y} copies is successfully received at the sink, which is defined as the \textit{QoS} delivery probability of the node. The \textit{$Q(y_c)$}, the collision transmission of delivery probability of each packet and is given by 

\begin{equation}
Q(y_c) = (1 - Q_{nc})^y
\end{equation}

The probability of successful transmission of sensor data packet \textit{$Q(y_s)$} is given by 

\begin{equation}
Q(y_s) = (1 - Q(y_c))
\end{equation}

Combining the above equations, we have

\begin{equation}
Q(y_s) = (1 - Q(y_c))  \nonumber \\
\end{equation}

\begin{equation}
Q(y_s) = 1 - (1 - Q_{nc})^y   \nonumber \\
\end{equation}

\begin{equation}
Q(y) = Q(y_s) = 1 - (1 - e^{-2\beta T_{p}})^y
\end{equation}

The \textit{Q(y)} expresses the \textit{QoS} delivery probability as function of the number of retransmissions attempted by each node in the interval \textit{$T_{p}$}. 

The maximum delivery probability \textit{$Q_{max}$} that can be achieved is given by 

\begin{equation}
Q_{max} = 1 - (1 - e^{-2\beta T_{p}})^y
\end{equation}

The above result gives relationship between the maximum delivery probability that can be achieved, the number of retransmission attempts that each node makes in every interval \textit{$T_{p}$} and the number of nodes \textit{$N$}.

\subsection{TWO-QoS GROUPS}
Consider delivery probability of two \textit{QoS} groups \textit{$C_{1}$} and \textit{$C_{2}$} containing \textit{$m_{1}$} and \textit{$m_{2}$} nodes and requiring minimum delivery probability \textit{$q_{1}$} and \textit{$q_{2}$} respectively. Number of nodes in \textit{$C_{k}$} is \textit{$m_{k}$} and each node in \textit{$C_{k}$} retransmits \textit{$y_{k}$} times in every interval \textit{$t_{k}$}. The number of retransmissions \textit{$y_{k}$} is the same for all the nodes in \textit{$C_{k}$}. 

The analysis is similar to that of one-hop retransmission, the probability of transmission from node in \textit{$C_{k}$} does not collide with transmission from any other node in the network and is given by

\begin{equation}
Q_{nc}(j) = e^{-2\beta_{tg}(\tau_{cs}+T_{p})}
\end{equation}

where, $\beta_{tg}$ is defined as the overall traffic generated by all nodes inside the transmission range of a node and its rate is given by 

\begin{equation}
\beta_{tg} = \sum_{k=1}^{k=2}\frac{m_{k} y_{k}}{t_{k}}
\end{equation}

The successful packet delivery probability achieved by node in \textit{$C_{k}$} can be expressed as 

\begin{equation}
Q_{suc}(k) = 1 - (1 - e^{-2\beta_{tg}(\tau_{cs}+T_{p})})^{y_{k}+1}
\end{equation}

where, $\tau_{cs}$ is the carrier sense period and $T_{p}$ is the duration of a packet transmission.

By the problem definition, $Q_{suc}(k) \geq q_{k}$, Then, we have

\begin{equation}
\label{eqn:9}
{1 - (1 - e^{-2\beta_{tg}(\tau_{cs}+T_{p})}) - (1 - q_{k})^ \frac{1}{(y_{k}+1)}} \indent \leq 0, k = 1 ... m_{k}
\end{equation}     \\              

\begin{equation}
\label{eqn:10}
{1 \leq y_{k} \leq \frac{t_{k}}{T_{p}}}, \indent k = 1 ... m_{k}
\end{equation}                   

The above mentioned constraint (9) guarantees that every node in group $C_{k}$ has the delivery probability of at least $q_{k}$ and constraint (10) states that the maximum number of retransmissions can not exceed $\frac{t_{k}}{T_{p}}$.

\section{Algorithm}         
In this paper, the performance of optimal retransmission algorithm \textit{GORMA} is discussed to find the solution to the optimization problem in one-hop \textit{QoS} group containing \textit{N} nodes. The objective of the \textit{GORMA} algorithm is to find the optimal retransmission value between $y_{low}$ and $y_{high}$ that minimizes the total sensor network traffic and each node in WSNs achieves maximum delivery probability in background traffic. 

Later, we have  two \textit{QoS} groups, consisting of $C_{1}$, $C_{2}$ containing $m_{1}$, $m_{2}$ nodes and  requiring minimum delivery probability of $q_{1}$, $q_{2}$ respectively. With two \textit{QoS} groups, we must have $y_{low} \leq y_{int} \leq y_{high}$.  Since $q_1 \ge q_2$, we need to have $y_{low} \leq y_{int}$. In addition, the nodes in $C_{2}$ achieve delivery probability of at least $q_{2}$. The background network  traffic for node in $C_{2}$ must be bounded by 
      
\begin{equation}
\label{eqn:11}
(m_{1} y_{low} + m_{2} y_{int}) \leq (m_{1} y_{high} + m_{2} y_{high}) 
\end{equation}                   

which implies that
     
\begin{equation}
\label{eqn:12}
y_{low} \leq y_{high} +  \frac{m_{2}(y_{high} - y_{int})}{m_{1}}
\end{equation}                   
            
From the above discussion, the binding values are $y_{low}$ and $y_{high}$, we need to pass through $y_{int}$ values from $y_{low}$ to $y_{high}$ to satisfy minimum delivery probability $q_{1}$ and $q_{2}$. Otherwise, the algorithm declares that no feasible solution exists.    

\textit{GORMA} algorithm solves the problem of energy consumption and delivery probability.

The sensor nodes are randomly deployed and sends data packet to a sink node, \textit{y} number of times. All the \textit{WSNs} nodes are within one hop transmission range of the sink and source nodes in the network to achieve the same delivery probability as shown in Algorithm 1. When a data packet is received by a node after transmitting \textit{y} number of times, in a given period. The arrival rate of packet is modeled as a poisson distribution and the maximum delivery probability is achieved in one hop retransmission. The \textit{N} nodes are divided into Two-QoS groups $C_1$ and $C_2$ consisting of $m_1$ and $m_2$ nodes. The minimum delivery probability of each group are $q_1$ and $q_2$. In Two-QoS groups, we must have $y_{low} \leq y_{int} \leq y_{high}$.  Since $q_1 \ge q_2$, we need to have $y_{low} \leq y_{int}$. It minimizes the total network traffic and improves the delivery probability. 

\begin{algorithm}
$begin$ \\
All $N$ Sensor Nodes are within the Sensing and Communication Range  \\
Nodes are Randomly Distributed \\
All Source Nodes Send Data Packet to a Sink Node \\
Sink has information about each Source Node Location and \textit{ID} \\
Each Node Energy Depends on Distance and Data Size \\
Each Source Nodes are Transmits $y$ copies at Random Instant \\
Data Packets are received within the given time Period at Sink \\
Delay from each Source Node to a Sink is same \\
Probability of Error is Minimized \\
$N$ Sensor Nodes are Divided into Two \textit{QoS} Groups $C_1$ and $C_2$\\
$m_1$ and $m_2$ Sensor Nodes are Deployed in Each \textit{QoS} Group \\
$q_1$ and $q_2$ are Minimum Delivery Probability \\
$y_k$ Times Retransmits in each \textit{QoS} Group \\
Minimize Total Network Traffic \\
Minimum Number of Retransmission from each \textit{QoS} Group \\

$Q(y) \gets 1 - (1 - e^{-2\beta T_{p}})^y$ \\

$y_{int} \gets y_{low}$\;
$y_{high} \gets y_{high} +  \frac{m_{2}(y_{high} - y_{int})}{m_{1}}$\;

\For{$y_k \gets y_{int} \hspace{0.05in}  \textbf{to} \hspace{0.05in}  y_{high}$} {
  $Q_{suc}(k) = 1 - (1 - e^{-2\beta_{tg}(\tau_{cs}+T_{p})})^{y_{k}+1}$ \\
   \If{$Q_1(y_1) > q_1$} {
    $Success \gets True$\;
  }
   \Else {
    No feasible solution exits
  }
}
\Return{$y$}\;
\caption{{\sc GORMA} Algorithm}
$end$ \\
\end{algorithm}

\section{PERFORMANCE EVALUATION}
\subsection{Simulation Setup}
The performance of \textit{GORMA} has been evaluated using \textit{NS2} simulator package to obtain packet delivery probability and energy consumption. A random flat-grid scenario is chosen for deployment of the nodes within 50m x 50m and 230m x 230m area. In our simulation model, we use two-ray ground reflection model for radio propagation and omnidirectional antenna. The transmission bandwidth is set to 50 Mbps, each source node has only transmit circuit and no receiver. For \textit{One-hop} retransmission and \textit{Two-QoS} groups the number of nodes \textit{N} is 100, data arrival rate \textit{T} = 1ms and packet transmission time $T_{p}$ = 6.4 x $10^{-4}$ ms.

\begin{figure}
\begin{center} 
\includegraphics[width=20pc,height=14pc]{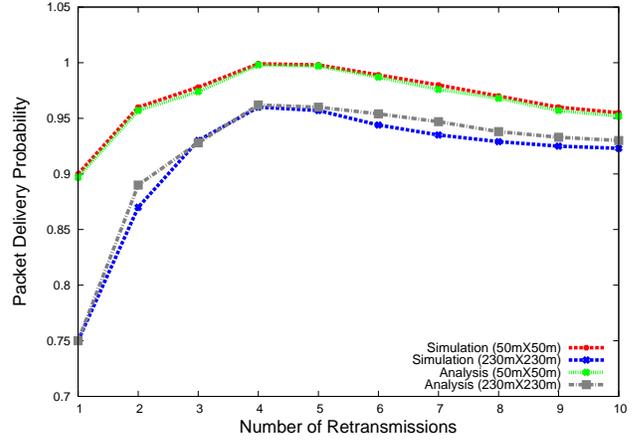}
\caption{Analysis and Simulation results for the One-hop QoS group}
\end{center}
\end{figure} 

\subsection{One-Hop QoS Group}
The results for \textit{Q(y)} given in equation 4 for one-hop retransmission, consisting of \textit{N} = 100 nodes is plotted in Figure 2. It shows that the probability of the delivery of packets initially increases with the number of retransmissions, reaches maximum and then decreases. The simulation and numerical analysis results shows that the maximum delivery probability of \textit{Q(y)} is 0.9990 for 50m x 50m area when y = 4 or y = 5. The mimimum delivery probability \textit{Q(y)}is 0.978 is achieved for $3 \leq y \leq 9$. The \textit{GORMA} scheme minimizes the network traffic when y = 3 and maximizes the probability of delivery of data packets when the retransmission value y = 4. 

The second set of curves of \textit{GORMA} of simulation results is comparable with the theoretical analysis. The delivery probality \textit{Q(y)} is 0.96 when the nodes are randomly distributed in 230m x 230m region. Since, the simulation performance of sensor nodes are poor in a large region, we assume that the packet loss is only due to channel errors and not due to collisions or interference.   

The graph in Figure 2 illustrates that the number of retransmission by each sensor node is reduced by choosing the value for y as 3 or 4. This increases the probability of delivery, which in turn increases the lifetime of the sensor nodes. The simulation results are within the scopes of the analytical results. 

\begin{figure}
\begin{center} 
\includegraphics[width=20pc,height=14pc]{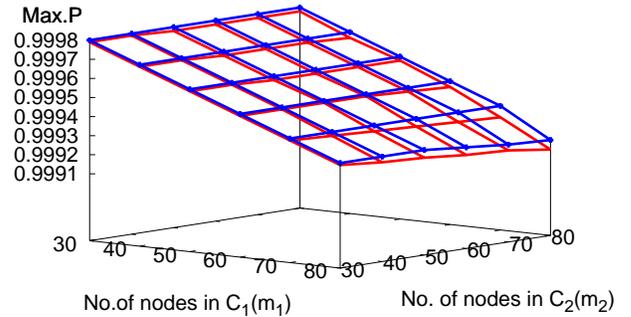}
\caption{Analysis and Simulation results of Max.P for Two QoS Groups}
\end{center}
\end{figure} 

\begin{figure}
\begin{center} 
\includegraphics[width=20pc,height=14pc]{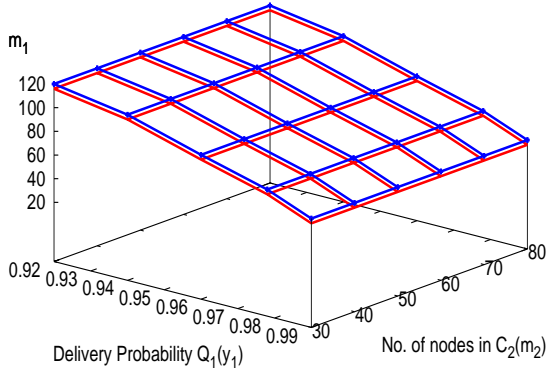}
\caption{Analysis and Simulation results for Two QoS Groups}
\end{center}
\end{figure} 

\subsection{Two QoS Groups}
For analysis and simulation, we assume that there are only two \textit{QoS} groups. The delivery probability of two \textit{QoS} groups are $C_1$ and $C_2$ consisting of $m_1$ and $m_2$ nodes and requiring minimum delivery probability $q_1$ and $q_2$ respectively. 

Figure 3 shows the simulation and numerical analysis for solutions to the optimization problem for maximizing the delivery probability. The delivery probability $Q_{1}(y_{1})$ is high for small size of networks, say, for available nodes $m_1$ = 30 and $m_2$ = 30. Then, it is possible to achieve $Q_{1}(y_{1})$ is 0.9999. For $m_2$ = 80 and small value of $m_1$ = 30, $Q_{1}(y_{1})$ is 0.9998. Similarly, with large number of nodes $m_1$ = $m_2$ = 80, the delivery probability $Q_{1}(y_{1})$ drops to nearly 0.9996. This confirms that the delivery probability $Q_{1}(y_{1})$ depends on the number of sensor nodes in each QoS groups. 

Figure 4 shows the values of maximum achievable $m_1$ when $Q_{1}(y_{1})$ and $m_2$ are given. Assuming that $Q_{2}(y_{2})$ is 0.9, 120 nodes can achieve the required delivery probability $Q_{1}(y_{1})$ = 0.92, when $m_2$ node is 30. Similarly, when $Q_{1}(y_{1})$ = 0.99 and $m_2$ is 30, for this the maximum achievable nodes $m_1$ = 62.

\subsection{MINIMIZING ENERGY CONSUMPTION}
\begin{figure}
\begin{center} 
\includegraphics[width=20pc,height=14pc]{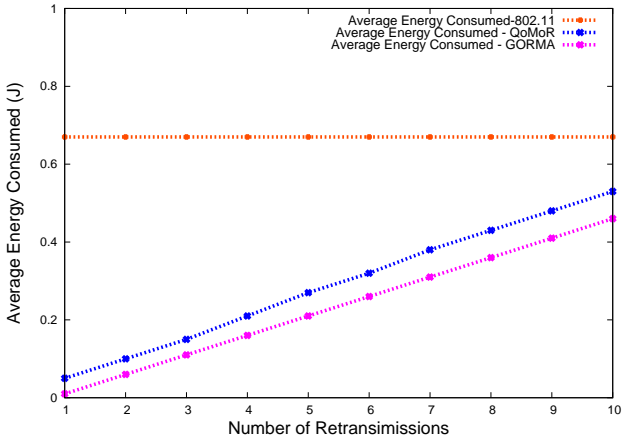}
\caption{Average Energy Consumption}
\end{center}
\end{figure} 

\begin{figure}
\begin{center} 
\includegraphics[width=20pc,height=14pc]{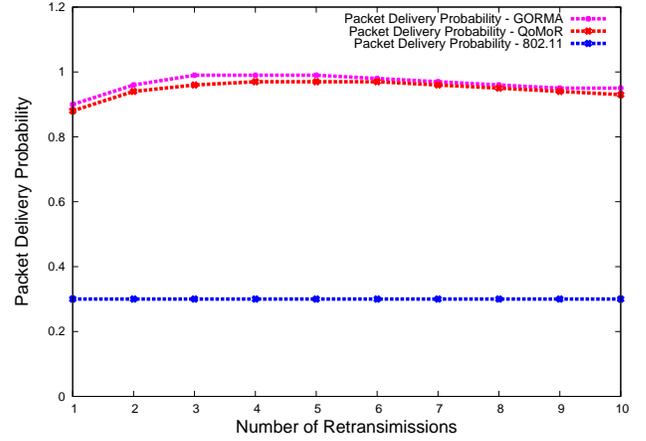}
\caption{Packet Delivery Probability}
\end{center}
\end{figure} 

The goal in \textit{GORMA} protocol is to minimize the energy consumption between any source nodes and sink node. The lifetime of sensor nodes in the network is directly proportional to the energy dissipation of each sensor node. The consumed energy in sensors includes the energy required for sensing, transmitting, receiving and processing of data. \textit{GORMA} protocol contribute to energy efficiency by minimizing collisions and retransmissions. We simulate the performance of \textit{GORMA} protocol respect to energy consumption and compare the average energy consumption with \textit{QoMoR} and \textit{802.11b}. 

The goal of the Collision Avoidance protocol is based on Medium Access Control of the \textit{GORMA} to increase the channel access probability for fairly distributing the energy consumption of the stations and thereby increasing the network lifetime.  Figure 5 shows the average energy consumption of the \textit{GORMA} scheme for different values of retransmissions about 2K data packets when the aggregated data rate generated by all the nodes is about 50Mbps, which is equal to the available bandwidth. The energy consumed by the \textit{GORMA} scheme for the number of retransmissions value 10, is less than the energy consumed by the \textit{QoMoR} and \textit{802.11b} protocol. The \textit{GORMA} protocol uses shorter frame slots, avoiding control packets like \textit{RTS} and \textit{CTS}, which unnecessarily consume energy and bandwidth. 

Figure 6 shows the delivery probabilities achieved by \textit{GORMA} protocol, \textit{QoMoR} and \textit{802.11b}  under the same conditions. The \textit{GORMA} protocol is significantly higher than that achieved by the \textit{QoMoR} and \textit{802.11b} protocol. Both \textit{QoMoR} and \textit{802.11b} does not use the available bandwidth as efficiently as \textit{GORMA}. The \textit{GORMA} provides QoS to the nodes using random access mode where each node transmits data packet by selecting variable slots randomly for variable data packet according to the total number of sensors nodes and each nodes take local decisions, depending on some pre-defined efficient retransmission probabilities.  

When the number of nodes is large and the aggregate data rate is matching to the available channel bandwidth, the performance of the \textit{GORMA} protocol is significantly better than \textit{QoMoR} and \textit{802.11b} both in terms of \textit{QoS}, delivery probability and energy consumption for the event-driven applications.

\section{CONCLUSIONS}
In this paper, we have implemented a Group Based Optimal Retransmission Medium Access Protocol (\textit{GORMA}), which combines \textit{CA} and efficient energy management protocol for low-cost, short radio range and low-energy \textit{WSNs} applications like Smart Environment, Home Automation, Structure Monitoring, Intelligent Transportation and Medical Monitoring. In this protocol, we have assumed that the source nodes do not have receiver circuits. Hence, they can only transmit data packet to sink node, but cannot receive any control signals from sink. In our work, we have designed a mathematical model to evaluate the maximum delivery probability and minimize the energy consumption by optimal retransmission technique. In proposed protocol \textit{GORMA}, each source node simply retransmits each of its data packet an optimal number of times within a given period of time in one-hop \textit{QoS} group and two \textit{QoS} groups. The simulation results show that, each source nodes employ probabilistic retransmission to minimize the energy consumption and maximize the packet delivery probability. 

 \bibliographystyle{IEEEtran}
 \bibliography{Bibliography}
\begin{biography}[{\includegraphics[width=1in,height=1.25in,clip,keepaspectratio]{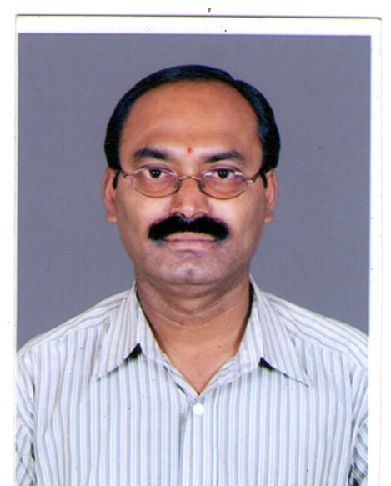}}]{Kumaraswamy M}
received B.E. degree in Electrical and Electronics Engineering from NIE, University of Mysore, Mysore. He obtained M.Tech in System Analysis and Computer Applications from NITK Surathkal. He is presently pursuing his Ph.D programme in the area of Wireless Sensor Networks in JNTU Hyderabad. His research interest includes Wireless Sensor Networks and Adhoc Networks. 
\end{biography}
\begin{biography}[{\includegraphics[width=1in,height=1.25in,clip,keepaspectratio]{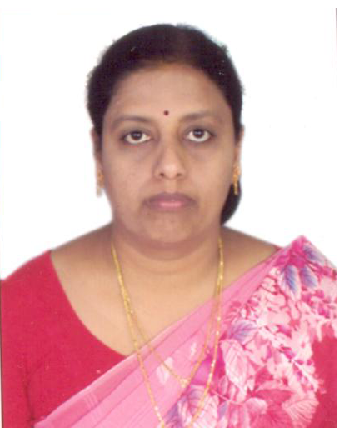}}]{Shaila K}
is Professor and Head of the Department of Electronics and Communication Engineering at Vivekananda Institute of Technology, Bangalore, India. She obtained her B.E and M.E degrees in Electronics and Communication Engineering from Bangalore University, Bangalore. She was awarded Ph.D programme in the area of Wireless Sensor Networks in Bangalore University, Bangalore. She has published a book on Digital Circuits and System. Her research interest is in the area of Sensor Networks, Adhoc Networks and Image Processing.
\end{biography}
\begin{biography}[{\includegraphics[width=1in,height=1.25in,clip,keepaspectratio]{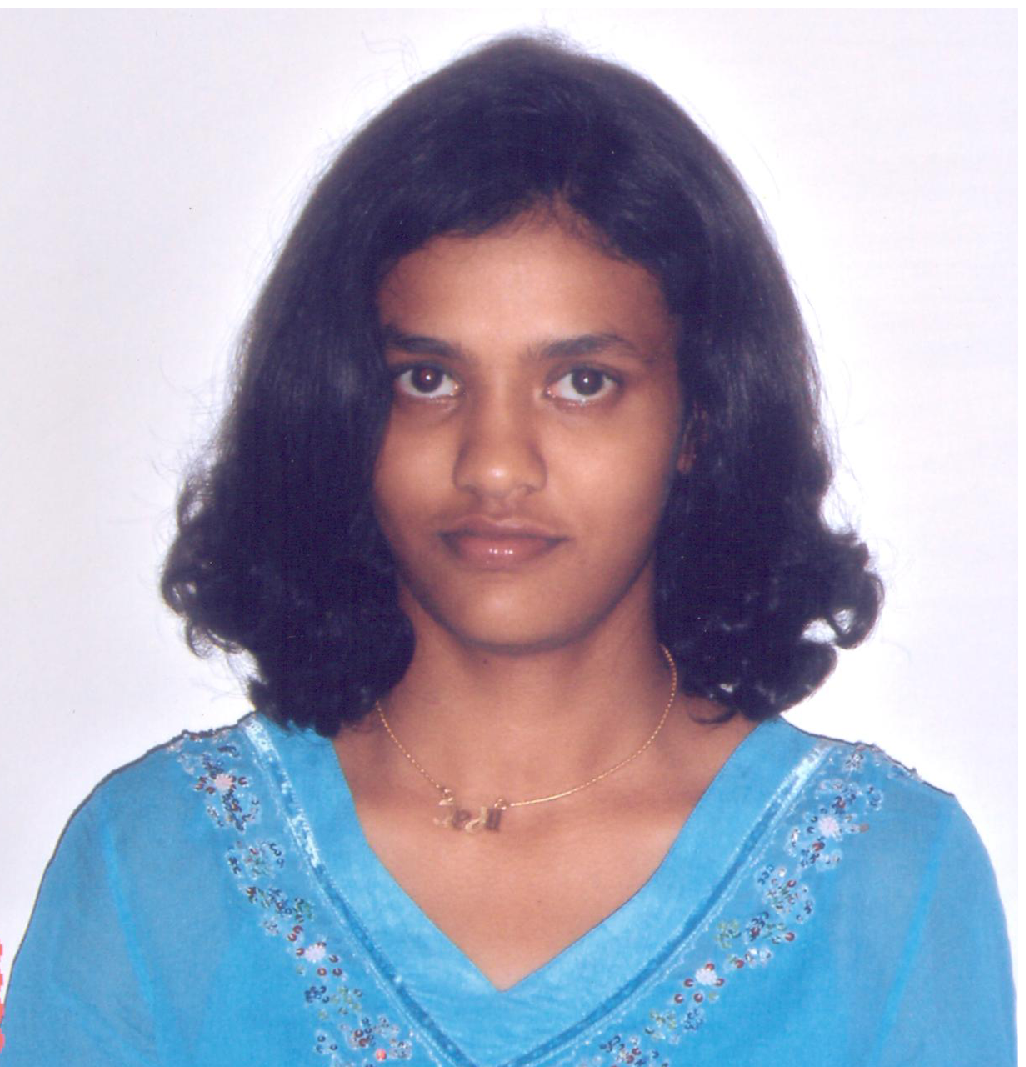}}]{Tejaswi V} 
completed her graduation in CSE from Rastriya Vidayala College of Engineering, Bangalore. She is currently pursuing her M.Tech (CSE) from NITK Surathkal. Her research interest is in the area of Wireless Sensor Networks.
\end{biography}
\begin{biography}[{\includegraphics[width=1in,height=1.25in,clip,keepaspectratio]{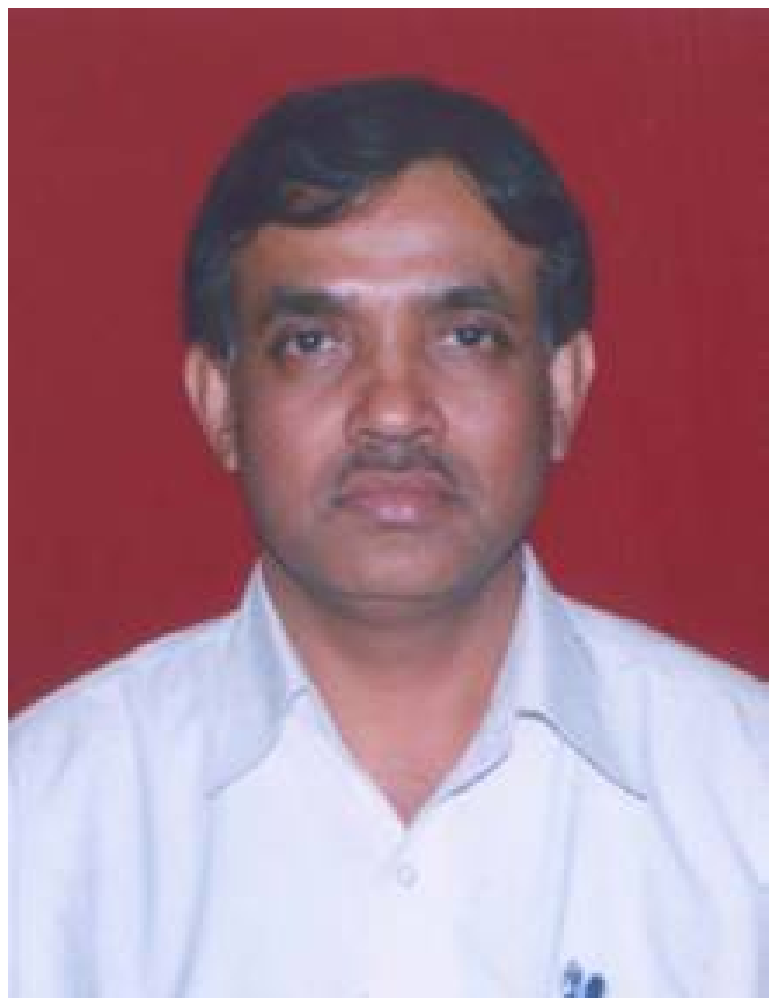}}]{Venugopal K R}
is currently the Principal, University Visvesvaraya College of Engineering, Bangalore University, Bangalore. He obtained his Bachelor of Engineering from University Visvesvaraya College of Engineering. He received his Masters degree in Computer Science and Automation from Indian Institute of Science Bangalore. He was awarded Ph.D. in Economics from Bangalore University and Ph.D. in Computer Science from Indian Institute of Technology, Madras. He has a distinguished academic career and has degrees in Electronics, Economics, Law, Business Finance, Public Relations, Communications, Industrial Relations, Computer Science and Journalism. He was a Postdoctoral research scholar in University of Southern California, USA. He has authored and edited 39 books on Computer Science and Economics, which include Petrodollar and the World Economy, C Aptitude, Mastering C, Microprocessor Programming, Mastering C++ and Digital Circuits and Systems etc.. During his three decades of service at UVCE he has over 400 research papers to his credit. His research interests include Computer Networks, Wireless Sensor Networks, Parallel and Distributed Systems, Digital Signal Processing and Data Mining.
\end{biography}
\begin{biography}[{\includegraphics[width=1in,height=1.25in,clip,keepaspectratio]{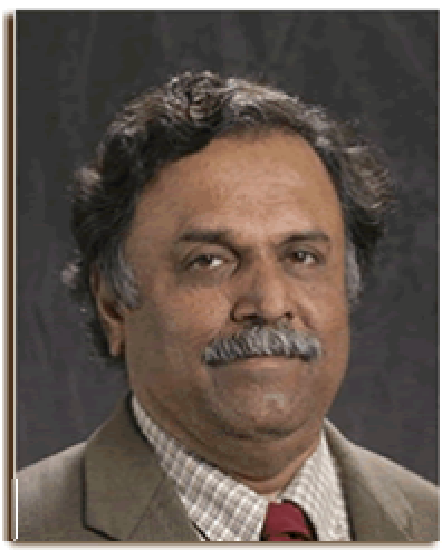}}]{S S Iyengar}
is currently the Roy Paul Daniels Professor and Chairman of the Computer Science Department at Louisiana State University. He heads the Wireless Sensor Networks  Laboratory and the Robotics Research Laboratory at LSU. He has been involved with research in High Performance Algorithms, Data Structures, Sensor Fusion and Intelligent Systems, since receiving his Ph.D degree in 1974 from MSU, USA. He is Fellow of IEEE and ACM. He has directed over 55 Ph.D students and 100 Post Graduate students, many of whom are faculty at Major Universities worldwide or Scientists or Engineers at National Labs Industries around the world. He has published more than 500 research papers and has co-authored 6 books and edited 7 books. His books are published by John Wiley and Sons, CRC Press, Prentice Hall, Springer Verlag, IEEE Computer Society Press etc.. One of his books titled Introduction to Parallel Algorithms has been translated to Chinese. 
\end{biography}
\begin{biography}[{\includegraphics[width=1in,height=1.25in,clip,keepaspectratio]{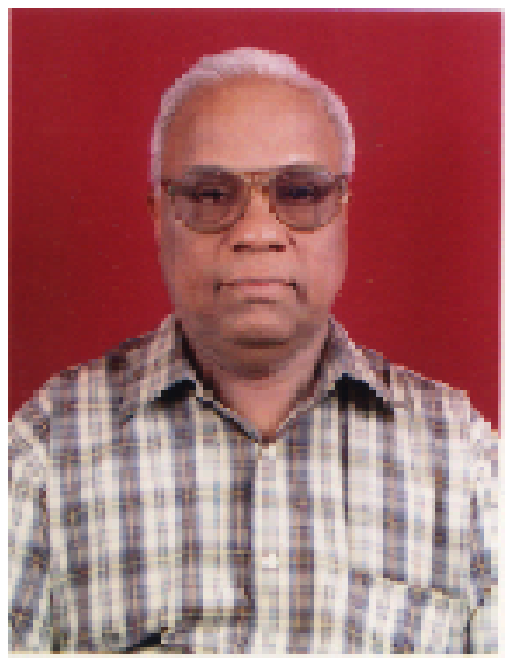}}]{L M Patnaik}
is an Honorary Professor, Indian Institute of Science, Bangalore, India. He was a Vice Chancellor, Defense Institute of Advanced Technology, Pune, India. He was a Professor since 1986 with the Department of Computer Science and Automation, Indian Institute of Science, Bangalore. During the past 35 years of his service at the Institute he has over 700 research publications in refereed International Journals and refereed International Conference Proceedings. He is a Fellow of all the four leading Science and Engineering Academies in India;  Fellow of the IEEE and the Academy of Science for the Developing World. He has received twenty national and international awards; notable among them is the IEEE Technical Achievement Award for his significant  contributions to High Performance Computing and Soft Computing. His areas of research interest have been Parallel and Distributed Computing, Mobile Computing, CAD for VLSI circuits, Soft Computing and Computational Neuroscience.
\end{biography}

\end{document}